\documentstyle[12pt,multicol,amsfonts,epsf]{article}

\newcommand{\cD}{{\cal D}}

\newcommand{\beq}{\begin{equation}}
\newcommand{\eeq}{\end{equation}}
\newcommand{\beqy}{\begin{eqnarray}}
\newcommand{\eeqy}{\end{eqnarray}}

\newtheorem{Definition}{Definition}
\newtheorem{Lemma}{Lemma}
\newtheorem{Theorem}{Theorem}

\newenvironment{Definition*}{{\bf Definition}}{}

\def\C{{\mathbb{C}}}
\def\Z{{\mathbb{Z}}}
\def\R{{\mathbb{R}}}

\def\N{{\mathbb{N}}}

\newcommand{\cH}{{\cal H}}

\date{{\small Sept 30, 2004}}

\title{\Large \textbf{Decomposition 
of time-covariant operations on quantum systems with
continuous and/or discrete energy spectrum}}

\author{Dominik
Janzing\thanks{e-mail: 
janzing@ira.uka.de} \\ 
{\small IAKS Prof.~Beth, Arbeitsgruppe Quantum Computing,}\\ 
{\small Universit{\"a}t Karlsruhe,}\\{\small Am Fasanengarten 5,
D-76\,131 Karlsruhe, Germany}}

\begin{document}
\maketitle

\begin{abstract}  
Every completely positive map $G$ that commutes which the Hamiltonian
time evolution is  an integral or sum over 
(densely defined) CP-maps $G_\sigma$ 
where $\sigma$ is the energy that is transferred to or taken from the
environment.
If the spectrum is non-degenerated each $G_\sigma$ is a 
{\it dephasing} channel
followed by 
an {\it energy shift}.
The dephasing is given by the Hadamard 
product of the density operator
with a (formally defined) positive operator.
The Kraus operator of
the energy shift is a partial isometry which defines a translation
on $\R$ with respect to a non-translation-invariant measure. 

As an example, I calculate this decomposition explicitly for the
rotation invariant gaussian channel on a single mode.

I address the question under what conditions a covariant channel
destroys superpositions between mutually orthogonal states
on the same orbit. 
For  channels which allow mutually orthogonal output states
on the same orbit,
a lower bound on 
the quantum capacity is derived
using the
Fourier transform of the CP-map-valued measure $(G_\sigma)$. 

\end{abstract}

\section{Introduction}

It was an important insight in the early days of 
quantum computing and quantum cryptography research
that quantum theory should not be considered  as a purely physical theory, 
but it
rather defines a new kind of information, called 
quantum information. The decisive feature is its fragility since
quantum information 
is destroyed whenever one tries  to copy it \cite{NoCloning}. 

The question which physical processes and information channels
preserve quantum information plays therefore a crucial role in the theory
of quantum communication. A central concept for addressing this issue
is the notion of a {\it channel}. Here we 
consider a channel  or (``operation'') to be a
completely positive trace preserving map $\rho \mapsto G(\rho)$ where
$\rho$ and $G(\rho)$ are positive operators of trace one acting
on Hilbert spaces $\cH_{in}$ and $\cH_{out}$, respectively \cite{Kraus}.
Analyzing quantum or classical channel capacities is in general 
a difficult task. Here we consider a certain
 kind of channels or operations,
namely those which are {\it time-covariant}.
For simplicity,
we will assume that the input and the output state is a state
of the same physical system. Then we have $\cH_{in}=\cH_{out}=:\cH$.
Furthermore we have a Hamiltonian time evolution on $\cH$ generated by
a densely defined self-adjoint operator $H$ which reads
\begin{equation}\label{Cov1}
\alpha_t (\rho):=e^{-iHt} \,\rho \,e^{iHt}\,,
\end{equation}
where $\rho$ is a positive operator with trace one. 
We call a channel $G$ time-covariant\footnote{Note that
in the context of classical linear systems
the corresponding property is  usually referred to 
time-``invariance''.} if it satisfies the equation 
\begin{equation}\label{cov}
\alpha_t \circ G =G \circ \alpha_t\,.
\end{equation}
It is a special case of the requirement
\[
G(U_g \rho U_g^\dagger)=U_g G(\rho) U_g^\dagger\,,
\]
where $g\mapsto U_g$ is an arbitrary unitary group representation. 

For compact groups, each 
covariant CP-map $G$ has
a representation (see \cite{HolevoClass} which refers partly to \cite{Barut})
\[
G(\rho)=\sum_j L_j\rho L_j^\dagger
\]
where the Kraus operators $L_j$ satisfy 
\begin{equation}\label{Ucov}
U_g L_j U_g^\dagger =\sum_k d_{jk}(g) L_k\,,
\end{equation}
and  $g\mapsto [d_{jk}(g)]_{j,k\leq d}$ is a $d$-dimensional 
unitary representation with arbitrary $d$ (in
\cite{HolevoClass} it is mainly focussed on the case that
$g \mapsto U_g$ is an irreducible representation).
Note that the time evolution can only be described
by a compact group if it is periodic which cannot be guaranteed
even in the finite dimensional case.
In \cite{HolevoMark}
the case is considered that
$G$ is replaced by a whole dynamical semi-group $(G_t)_{t\in \R^+}$ 
which is covariant
with respect to a group representation.

Condition (\ref{cov}) 
appears naturally, for instance, in the following situations:

\begin{enumerate}

\item {\bf Theory of timing information:} 
 Assume that the density matrix $\rho$ is subjected to an arbitrary 
quantum operation $G$ at  
a completely 
unknown time instant. Then the statistical description of this operation
leads to a map $G'$ which satisfies necessarily 
(\ref{cov}). This is the key idea
in \cite{clock} where condition (\ref{cov}) defines a ``quasi-order
of clocks'' which
 classifies systems with respect 
to their timing information. The time-covariant maps are exactly those
which can be implemented without using external clocks.
Given a non-stationary state $\rho$, the
set of states which can be obtained from $\rho$ using time-covariant 
operations  are  those which have at most as much timing information
as $\rho$.

\item {\bf Decoherence:} Dephasing of systems is described by a decay 
of the off-diagonal entries
with respect to the energy eigenbasis.  This channels satisfy clearly 
condition (\ref{cov}).

\item {\bf Scattering processes / Quantum generalization of 
transfer functions:} 
Time-covariant operations appear naturally in the description of 
scattering processes. Then $G$ generalizes the scattering 
operator by including classical or quantum stochastic fluctuations.
Here  {\it scattering} is understood in a rather general sense. 
Apart from the situation that a particle is scattered by the potential
of another particle, one may, for instance, also think of a
 light beam  that passes a filter.
One may consider  time-covariant maps as
quantum generalization of classical time-invariant linear devices which
are described by their transfer functions.
\end{enumerate}
 
The paper is organized as follows. 
In Section \ref{Transfer} we recall briefly
how to describe
classical time-invariant linear systems in signal processing
by a transfer function \cite{ZadehLinear}. 
To figure out to what extent this concept
can be generalized to the quantum stochastic 
setting is the key motivation of this paper.
In Section \ref{Scat} we recall abstract scattering theory in Hilbert spaces
and explain in Section \ref{TimeIn} why  time-covariant channels
are also considered 
as a natural generalization of scattering theory. In Section \ref{Had} we consider CP-maps
which are given by the Hadamard product of the density matrix with a
positive matrix. These maps will turn out as a building block of our
decomposition.
In Section \ref{Form} we derive the general form of a time-covariant 
CP-map.
The main result is that every time-covariant CP-map 
acting on a system with non-degenerate Hamiltonian
has a (densely defined)
decomposition 
as an integral or sum over 
(densely defined)
CP-maps $G_\sigma$ is a dephasing followed by an energy shift.
In Section \ref{Bounds} we address the question whether and to what extent
covariant channels can destroy superpositions between states
on the same orbit (with respect to the time-evolution).
We use the general decomposition
to derive a lower bound on the quantum capacity 
of a specific type of time-covariant channels, namely those 
which have as output mutually orthogonal density operators on the same orbit.
In Section \ref{Gauss} we apply the decomposition to a single
mode gaussian channel.

\section{Classical time-invariant linear systems}

\label{Transfer}

To show the differences and the analogies to the theory 
of classical linear time-invariant channels
we briefly rephrase the concept of transfer functions
which is among the most important tools in
classical signal processing.

Consider a classical channel with an incoming and an outgoing signal.
These may, for instance, be light pulses or electrical pulses,
acoustical or other signals.
Let $t\mapsto Y(t)$ be the incoming signal where $Y(t)$ is the value
of an arbitrary physical quantity at the time instant $t$.
Let $Y'(t)$ denote the outgoing signal. 
In order to avoid problems with undefined Fourier transforms
we assume that $Y$ and $Y'$ are vectors in $L^2(\R)$, the Hilbert space of 
square-integrable functions on $\R$.
We assume that the system is described by a linear bounded operator 
$A$ on $L^2(\R)$ with
\[
Y'=AY\,.
\]
Assuming  time-invariance of the channel we have
$
Y'(t+s)= (A Y)(t+s)\,.
$ 
Defining the group of shifts $(S_t)_{t\in \R}$ by
\[
(S_tY)(s)=Y(t-s)
\]
we have
\[
S_t A =A S_t\, \,\, \forall t\in \R\,.
\] 
The implications of this condition can easily be derived 
by applying Fourier transformation to both sides of the equation:
The shift operators $S_t$ act then as the multiplication operators $M_t$ 
\[
\hat{Y} \mapsto M_t \hat{Y}
\]
with 
\[
(M_t\hat{Y})(\omega)= e^{i\omega t} \hat{Y}(\omega)\,.
\]
A linear operator commuting with all operators $M_t$ 
is necessarily a  
multiplication operator as well (see \cite{Baumgaertel}, 12.1.5). 
Therefore $A$ can be characterized by
a so-called transfer function $a$ with
\[
\hat{Y'}(\omega)= a(\omega) \hat{Y}(\omega)\,.
\]
Note that it is essential that the quantity $Y(t)$ is a scalar. 
If $Y(t)$ is a vector of dimension $d$  greater than $1$ the spectrum
of the time-evolution group $S_t$ is degenerated and
$\hat{Y}(\omega)$ is an element of a vector space $\C^d$ or $\R^d$ and
the transfer function $a$ had to be replaced by a $d\times d$-matrix. 
It is straightforward to ask whether time-covariant quantum operations
allow a natural generalization of the transfer function.
However, one should recall that the situation with non-degenerated 
spectrum is possible on the Hilbert space level (this is well-known
in abstract scattering theory, as will be rephrased in the next section) 
but not on the level of
density matrices, where the dynamics is generated by the 
super-operator $-i[H,.]$. The latter has always degenerated spectrum.

\section{Abstract scattering theory}

\label{Scat}
A similar approach as above applies to a scattering process of a quantum
mechanical particle: A particle which comes from the infinity
and passes a scattering potential. After it has left the potential 
it disappears to infinity. Here we only recall some
standard results of scattering theory \cite{Yafaev,Baumgaertel}
which are essential for this paper.
Let $H=H_i+H_0$ be the total Hamiltonian of the system which consists
of the free Hamiltonian $H_0$ and the interactions term $H_i$.
One assumes that the particle moves approximatively 
according to its free Hamiltonian $H_0$ 
for $t \to \pm \infty$. 
For  potentials which  decay sufficiently with the distance
between particle and the scattering center,
the limits
\[
|\psi^\pm\rangle:= \lim_{t\to 
\pm\infty} \exp(iH t) \exp(-iH_0t) |\psi\rangle 
\]
exist \cite{Yafaev} on  an appropriate subspace
of wave functions $|\psi\rangle$.
There is a rich literature addressing the question 
under which circumstances and on which subspaces one can define
a unitary scattering operator\footnote{It is usual to denote the
scattering operator with $S$. On the other hand, it is 
usual to denote shifts by $S_t$ with some index $t$.
He hope that this will not lead to confusions.}
\[
|\psi^-\rangle \mapsto |\psi^+\rangle=:S|\psi_-\rangle\,.
\]
However, if it exists it commutes with the free Hamiltonian 
evolution \newline $\exp(-iH_0t)$. Therefore
the free time
evolution and the scattering operator $S$ can simultaneously be described by 
multiplication operators.
The analogy to Section \ref{Transfer} can even be closer in
a scattering process with  so-called Lax-Phillips evolution 
\cite{Baumgaertel}. 
Assume that there exist subspaces $\cH_-\leq \cH$ and $\cH_+\leq \cH$ 
(``incoming and outgoing subspaces'', respectively) such that 
\[
e^{-iHt} \cH_- \subset \cH_- 
\]
for all negative $t$ (the particle comes from the infinity) and 
\[
e^{-iHt} \cH_+ \subset \cH_+
\]
for all positive $t$ (the particle disappears to the infinity). 
Assume  furthermore 
that the intersection 
\[
\cap_{t\in \R} \exp(-iHt) \cH_\pm
\]
vanishes and that the span of all spaces $\exp(-iHt) \cH_+$
as well as the span of all $\exp(-iHt) \cH_-$
 is dense
in $\cH$.
Then 
one can assume without loss of generality that $\cH$ is the set of square 
integrable functions on $\R$ with respect to the Lebesgue measure, and 
$H$ is the multiplication operator $(H\psi)(\omega)=\omega \psi(\omega)$. 
The scattering operator is then (like the transfer function
in Section \ref{Transfer}) given by $(S\psi)(\omega) =s(\omega) \psi(\omega)$ 
with an appropriate function $s$.

\section{Quantum channels from scattering processes}

\label{TimeIn}
The unitary scattering operator
$S$ in Section \ref{Scat} 
defines a channel
$G(\rho)=S\rho S^\dagger$ which satisfies obviously the
time-covariance condition (\ref{cov}) 
with respect to the {\it free} evolution 
$\alpha_t(\rho)= \exp(-iH_0t)\rho \exp(iH_0t)$ 
due to $SH_0=H_0S$. It is clear that scattering with an unknown
potential could lead to an operation which is a statistical 
mixture of maps $S\rho S^\dagger$. But not only classical
fluctuations of the scattering potential lead to a mixture of
output wave functions; due to quantum fluctuations of the potential
one may also have a CP-map which is not a mixture of unitary scattering
processes. To see this,
consider a bipartite quantum system with Hilbert space
$\cH_A \otimes \cH_B$  that evolves according to
a joint Hamiltonian $H_{AB}$.
Assume that the joint evolution coincides
asymptotically with the separate evolution generated by $H_A +H_B$ 
and that the limits
\[
|\psi^\pm\rangle :=
\lim_{t\to \infty} \exp(itH_{AB}) \exp(-iH_At) \exp(-iH_Bt)|\psi \rangle
\] 
exist in an appropriate sense. Then we have formally
the same situation as in Section \ref{Scat} with the substitution
\[
H_0 \mapsto H_A+H_B\,,\,\,\,\,\,H\mapsto H_{AB}\,.
\]
If a scattering operator $U$ on  
$\cH_A \otimes \cH_B$ exists we have 
\begin{equation}\label{CommU}
[U,H_A+H_B]=0
\end{equation}
Let $\rho_A$  be an arbitrary initial density
matrix of system $A$. Then the scattering process defines a completely 
positive map on the system $B$ by
\[
G(\rho_B):=tr_A(U ( \rho_A\otimes \rho_B ) U^\dagger)\,.
\]
Due to (\ref{CommU})
it is easy to verify that $G$ is time-covariant if the initial state
$\rho_A$  is stationary with respect to the time evolution $\exp(iH_At)$.
Here we do not address the difficult question in which situation
the limits above exist. The problem of the existence of scattering operators
in the quantum stochastic setting is addressed in \cite{KuemmMaassen}.
This will not be our subject.
The remarks above were only to show that time-covariant channels
appear naturally also in the description of (possibly inelastic) scattering
processes.
Another situation where time-covariance appears is when 
the energy spectrum is discrete and a weak interaction with the environment
is switched on. Then the interaction implements clearly
a unitary $U$ on  system + environment such that $U$ satisfies
(\ref{CommU}).

\section{Quantum channels from Hadamard-products}

\label{Had}
Here we consider
a simple type of time-covariant CP-maps which will play a crucial role 
in the description of general time-covariant CP-maps.

The Hadamard product $A*B$ of two $n\times n$- matrices $A,B$ 
is defined as
 the entry-wise product $(A*B)_{ij}:=A_{ij}B_{ij}$.
Remarkably,  the Hadamard product between a density matrix $\rho$ 
with any positive
matrix $M$  of the same size $n$ defines a completely positive map:

If $M:=\sum_j r_j |d_j\rangle \langle d_j|$ is a spectral decomposition
of $M$  then we may define diagonal 
matrices $D_j$  which have the coefficients of
the vector $d_j\in \C^n$ as diagonal entries.
Then the map
\[
\rho\mapsto D_j \rho D_j^\dagger = (|d_j\rangle \langle d_j|) * \rho
\]
is obviously completely positive. So is the positive linear combination
of those maps.
Channels of this type have already be considered 
in \cite{HavelHad} to describe decoherence. It is clear that they
commute with the Hamiltonian time evolution if the Hadamard product
is calculated with respect to the energy basis.
In the finite dimensional case, it 
 we   obtain also time-covariant channels
by the following construction: Let $\Sigma:=\{\omega_1, 
\dots,\omega_n\}$ be the
eigenvalues of $H$ acting on $\C^n$. 
Let $|\omega_j \rangle$ be the corresponding eigenstates. For any $\sigma\in \R$ 
let $j_1,\dots,j_k$ be all indices $j$ such that $\omega_j+\sigma$ 
is in the spectrum $\Sigma$. Define the ``partial shift'' by
\[
S_\sigma:=\sum|\omega_{j_i} +\sigma \rangle  \langle \omega_{j_i}|\,,
\]
where the sum runs over all spectral values $\omega$ with
$\omega+\sigma \in \Sigma$  .

The map $\rho \mapsto S_\sigma \rho S_\sigma^\dagger$ is time-covariant
and also every map of the form
\[
\rho \mapsto \sum_\sigma S_\sigma (M_\sigma *\rho) S_\sigma^\dagger
\]
where the sum runs over all $\sigma$ for which a non-zero
partial shift exists. 
Each $M_\sigma$ is an arbitrary positive matrix. It is easy to check
that this map is trace-preserving if and only if 
\[
\sum_{\sigma} M_\sigma (\omega, \omega) =1\,\,\,\forall \omega
\in \Sigma\,,
\] 
where the sum runs over all values $\sigma$  for which there 
exist an $\omega \in \Sigma$ such that
$\omega+\sigma\in \Sigma$.

In the following we will show that this is the most general form
of time-covariant CP-maps. As will be shown, this holds
in principle even for the infinite dimensional case with
continuous spectrum when the sum is replaced by an integral
over a potentially uncountable number of energy shifts.

\section{The general form of time-covariant channels}

\label{Form}
To understand our construction it is helpful to consider the finite
dimensional  case first.
Using the Kraus representation \cite{Kraus}
\begin{equation}\label{Kr}
G(\rho)=\sum_j A_j \rho A_j^\dagger\,,
\end{equation}
one may choose $A_j$ such that they are eigenvectors of $\alpha_t$
(Since the derivation for the finite dimensional case
follows actually from the general derivation later in this section we only mention
briefly 
that this can be derived
as follows. 
One shows that
the representations  $t\mapsto [d_{kj}(t)]$ of $(\R,+)$
in eq.~(\ref{Ucov}) 
corresponding to time-translations can be choosen  
such that they are irreducible, i.e., one-dimensional).
The eigenvalue is 
 $\exp(-i\sigma_j t)$ where $\sigma_j$ may 
be any possible frequency difference
$\omega-\omega'$. For each possible value $\sigma$ we obtain
a  CP-map
\[
G_\sigma(\rho):=\sum_{\sigma=\sigma_j} A_j \rho A_j^\dagger\,.
\]
Let $A_j=U_j|A_j|$ be the polar decomposition of $A_j$ 
where the partial isometry 
$U_j$ is computed from the pseudo-inverse $|A_j|^{-1}$ by
\[
U_j:=A_j|A_j|^{-1}\,.
\]
One has  $\alpha_t(U_j)=\exp(-i\sigma_j t) U_j$ 
and  equivalently 
\begin{equation}\label{commrel}
[H,U_j]=\sigma_j U_j  \hbox{ and } [H,|A_j|]=0
\end{equation}
since $|A_j|$ is 
$\alpha_t$-invariant according to $\alpha_t(A_j)\alpha_t(A_j^\dagger)=
e^{-i\sigma_j t} e^{i\sigma_j t}$.
Using equation (\ref{commrel}) one checks easily that
\[
U_j |\omega\rangle = V_j |\omega+\sigma\rangle\, \,\,\forall \omega 
\in \Sigma
\]
where $V_j$ is an appropriate diagonal operator.
Hence we can write
\[
A_j = S_{\sigma_j} D_j
\]
with the diagonal matrix $D_j:=|A_j|V_j$.
Using the remarks of Section \ref{Had} we may write
\[
G_\sigma (\rho) = S_\sigma (M_\sigma *\rho) S_\sigma^\dagger
\]
where $M_\sigma=\sum_j |d_j \rangle \langle d_j|$ is defined as in Section 
\ref{Had} from the vectors $|d_j\rangle$ of diagonal entries of $D_j$.

We conclude:

\begin{Theorem}[Decomposition in finite dimensions]\label{FiniteMain}${}$\\
Let $G$ be a CP-map on $\C^n$ which commutes with the time evolution
$\alpha_t$ corresponding to a non-degenerate diagonal Hamiltonian.
Then $G$ has the form:
\[
G(\rho)=\sum_\sigma S_{\sigma}  (M_\sigma* \rho)  S^\dagger_\sigma\,,
\]
where the sum runs over all possible energy differences $\sigma$. Here
$S_\sigma$ denotes  partial shifts and $M_\sigma$ positive matrices.
\end{Theorem}

The map $\rho \mapsto (M_\sigma* \rho )$ preserves the energy of every state
and destroys to some extent the  coherent superpositions between them.
This is shown by the following two extreme cases:
$M_{\sigma}={\bf 1}$ where ${\bf 1}$ denotes the identity matrix. Then 
we have complete dephasing and obtain a mixture of energy eigenstates.
If $M_\sigma$ has only $1$ as entries it is the trivial channel which does
not affect the state at all.
The map $G$ consists of decohering channels followed by
different energy shifts.

To see the relation of our decomposition  
to the transfer function note
that all rank-one operators $|\omega+\sigma \rangle \langle \omega|$ 
span the eigenspace of $[H,.]$ with eigenvalue $\sigma$.
A time-covariant operation has to leave this subspace invariant.
Therefore one could decompose $G$ into a direct sum 
of linear maps $B_\sigma$ acting on these eigenspaces. However, 
in this decomposition one would have restrictions on the family $B_\sigma$ 
in order to yield a {\it completely positive map}.
Therefore we have  preferred
a decomposition where each component is completely positive.
However, the Hadamard multiplication for each frequency pair
$(\omega,\omega')$  
reminds to the multiplication
with the value of the transfer function for each frequency $\omega$. 

To generalize the decomposition to  infinite dimensions
we will need several vector-valued measures.
Therefore we recall the precise definition \cite{Diestel}:

\begin{Definition}{\bf (Vector-Valued Measures and Densities)}

\begin{enumerate}

\item
Let $\Omega$ be a set and $M$ the $\sigma$-algebra of measurable subsets
of $\Omega$.
For a Banach space $B$ 
a mapping $\gamma : M \rightarrow  B$ is called a  vector measure if 
\[
\gamma(\cup_j m_j)=\sum_j \gamma(m_j)
\]
for all finite collections of mutually disjoint sets $m_j$.
It is called countably additive when the same holds for countable sums.

\item
Let $\nu$ be a measure on the measure space $(\Omega,M)$ .
A measurable function $f:\Omega\rightarrow B$ is the Radon-Nikodym derivative
of $\gamma$ with respect to $\nu$  if
\[
\int_m f d\nu = \gamma (m) \,\,\,\,\forall m \in M\,.
\]
We also say that $f$ is the (vector-valued) density
of $\gamma$ with respect to $\nu$. 
\end{enumerate}
\end{Definition}

For trace-class operator-valued measures we will
have countable additivity in the weak sense, i.e.
that the scalar
measure that is given by evaluation on observables is
countably additive. For CP-map valued measures we demand 
countable additivity only after applying the maps to states and 
evaluating them on observables.

In order to construct the infinite dimensional analogue of $G_\sigma$ 
we
define
a function
\[
f_{K,\rho}(t):=tr(K G(\rho e^{-iHt})e^{iHt})\,.
\]
for every observable $K$ and state $\rho$.
First consider the simple case that $G=A\rho A^\dagger$ where $A$ is an operator
satisfying $\alpha_t(A)=e^{-i\sigma t}A$. Then 
we have
\[
f_{K,\rho}(t)=e^{i\sigma t} f_{K,\rho}(0)\,.
\]
If $G$ is defined by several Kraus operators $A_j$ which are eigenvectors
with different eigenvalues $\sigma_j$ the function $f_{K,\rho}$ would
consist of harmonic functions with all these frequencies $\sigma_j$.
Even though we do not expect in the infinite dimensional
case that we have Kraus operators which are eigenvectors of $\alpha_t$
it will turn out that we can nevertheless construct a decomposition
of $G$ based on the Fourier transform of $f_{K,\rho}$.
We show that
for each positive bounded operator $K$ and density operator $\rho$ 
the function $f_{K,\rho}$  is positive semidefinite, i.e., it satisfies
\[
\sum_{k,l} x_k \overline{x}_l f(t_k-t_l) \geq 0
\]
for all vectors $x\in \C^m$ and $m$-tuples $t_1,\dots,t_m \in \R$ with arbitrary $m$:
\begin{eqnarray*}
f_{K, \rho}  (t_k-t_l)&=&tr(K \sum_j A_j \rho e^{-iH(t_k-t_l)} A^\dagger_j
e^{-iH(t_l-t_k)})\\&=&\sum_j tr(K e^{iHt_k} A_j e^{-iHt_k} \rho
e^{iHt_l} A_j^\dagger e^{-iHt_l})\,,
\end{eqnarray*}
where we have used 
\[
\sum_j A_j \rho e^{-iHt_k} A_j^\dagger e^{iHt_k}=
\sum_j e^{iHt_k} A_j e^{-iHt_k}\rho A_j^\dagger\,,
\]
which is just another version 
of the time-invariance condition.
We have therefore
\begin{eqnarray*}
\sum_{k,l} x_k \overline{x}_l f(t_k-t_l)&=&
\sum_j tr (K (\sum_k x_k e^{iHt_k} A_j e^{-iHt_k})\rho (\sum_l \overline{x}_l e^{iHt_l}
A_j^\dagger e^{-iHt_l}))\\&=&
\sum_j tr(K C_j\rho C_j^\dagger) \geq 0
\end{eqnarray*}
with the abbreviation
\[
C_j:=\sum_k x_k e^{iHt_k} A_j e^{-iHt_k}\,.
\]
Since $f_{K,\rho}$ is positive semidefinite it defines a unique 
positive scalar 
measure $\nu_{K,\rho}$ on $\R$ by Bochner's theorem \cite{Knapp}.
For every measurable set $m\subset \R$ the
map $K \mapsto \nu_{K,\rho}(m)$ is a positive linear functional. 
It is norm-continuous (with respect to the operator norm) since
the norm of a positive functional  is given by
its value on the identity  \cite{Paulsen}, i.e., by $\nu_{{\bf 1},\rho}(m)$.
Now we restrict the functional to the set of compact operators
where
every norm-continuous functional  
is given by a trace-class operator \cite{Marchetti}. 
Therefore we may define a positive trace-class operator
$\rho_m$ by
\[
tr(\rho_m K) =\nu_{K,\rho} (m)\,.
\]
Hence the map $G_m$ defined by $G_m(\rho):=\rho_m$ 
is a positive map on the trace-class operators.
One can easily check that $G_m$
is also {\it completely} positive: one substitutes $G$ and $e^{-iHt}$ by 
an arbitrary tensor product extension $G\otimes id$ by
and $e^{-iHt}\otimes {\bf 1}$
and considers $\rho$ and $K$ as operators on the extended space. Then
it is obvious that the positivity argument above works similarly.
We have $G_\R=G$ due to
\[
tr(K \rho_\R )= \nu_{K,\rho} (\R) = f_{K,\rho} (0)=tr( K \rho ) \,.
\]
Since the trace of each $G(\rho)$ is at most $tr(G(\rho))=tr(\rho)=1$,
each map $G_m$ is a bounded operator on $T$, the set of trace-class operators.
Since $G_\R$ is trace-preserving the 
map $m \mapsto G_m$ is formally an {\it instrument} in the 
sense of Davies
(see \cite{Da78}, chapter 4). 
Think of $G_m(\rho)/tr(G_m(\rho))$ as the post-measurement
state given the knowledge that the measurement outcome $\sigma$ 
is in $m$. Then $\rho \mapsto G(\rho)$ is formally the 
effect of the measurement if the measured outcome is completely ignored.
Even though this interpretation refers to a virtual measurement
the virtual result $\sigma$  
has an observable interpretation:
Assume one observes the energy of the environment
before and after it has interacted with the system.
Due to energy conservation $\sigma$ is the energy loss of the environment.
Note that the initial energy of the environment can be observed 
without disturbing its state since we have assumed that it is 
in a stationary state.\footnote{The idea to observe the environment in
order to have a less mixed output state (which increases
information capacities) can already be found in \cite{Gregoratti,Hayden}.}
We shall call the probability measure $\nu_{\rho,{\bf 1}}$ 
the ``energy shift probabilities'' in the state $\rho$.

Each map $G_m$ is also time-covariant:
The obvious equation
\[
f_{K,\rho}=f_{\alpha_{t}(K),\alpha_t(\rho)}
\]
implies that it is irrelevant for the measure $\nu_{K,\rho}$ if
$\alpha_t$ is applied to  $\rho$ and $\alpha_{-t}$ to the output state
(which is equivalent to applying $\alpha_{t}$ to $K$). 

We summarize the results:

\begin{Theorem}
For each time-covariant channel $G$ 
there is a unique CP-map-valued-measure
\[
m\mapsto G_m
\]
such that 
\begin{enumerate}
\item 
Each $G_m$ is a time-covariant bounded operator on $T$.

\item The Fourier transform of $m\mapsto G_m$
is the function $t \mapsto \hat{G}_t$ with
\[
\hat{G}_t (\rho):=G(\rho e^{-iHt})e^{iHt}
\]
in the sense that 
\[
t\mapsto tr(K \hat{G}_t(\rho))
\]
is for all $K>0$ and all states $\rho$ 
the Fourier  transform of the (non-negative) measure 
\[
m \mapsto tr(K G_m(\rho))\,.
\]

\end{enumerate}

\end{Theorem}

The positive-map-valued measure defines a positive 
operator-valued-measure (POVM) \cite{Da78} $m\mapsto Q_m$ by
\[
tr(Q_m \rho):=tr(G_m(\rho))\,.
\]
In contrast to CP-map valued measures,
POVMs describe only the probabilities
for the measurement outcome without referring to the post-measurement state.
The Fourier transform of this POVM will play a crucial role later.
 
In order to have a stronger analogy to Theorem \ref{FiniteMain}
we would like to write $G$ as an integral
\[
G=\int G_\sigma d\nu(\sigma)
\]
with an appropriate measure $\nu$. Then the function 
$\sigma \mapsto G_\sigma$ would be
the CP-map valued Radon-Nikodym derivative of the measure
$m\mapsto G_m$, i.e., $G_\sigma$ would be the {\it density} of the measure
with respect to the measure $\nu$. 
The general 
problem of Radon-Nikodym derivatives of {\it instruments}
(CP-map valued measures) has already been considered in \cite{HolevoRa}.
It is shown that it exists in the following sense:

There is a $\sigma$-finite measure $\nu$, a dense domain $\cD\in \cH$,  
and a countable family of functions 
\[
\sigma \mapsto A_k^\sigma 
\]
defined for $\nu$-almost all $\sigma$ such that $A_k^\sigma$ are
linear operators $\cD \rightarrow \cH$ (not necessarily closable) such
that
\begin{equation}\label{HolevoBounded}
\int \sum_k   \|A_k^\sigma|\psi \rangle \|^2=\|\,|\psi \rangle \|^2\,\,\,\,\,\,\forall |\psi\rangle \in \cD
\end{equation}
and
\[
tr( K G_m(\rho))=\int_m \sum_k \langle A_k^\sigma \psi | K A_k^\sigma \psi \rangle \, d\nu (\sigma) \,\,\,\,\,\,\forall |\psi\rangle \in \cD\,.
\] 
Note that $A_k^\sigma$ could formally be considered as
the Kraus operators of a CP-map $G_\sigma$ with the ``only''
difference that Kraus operators are not only closable but even 
bounded.

Equation~(\ref{HolevoBounded})  states implicitly that 
\begin{equation}\label{HolSum}
 \sum_k \langle A_k^\sigma \psi | K A_k^\sigma \psi \rangle < \infty
\end{equation}
for $\nu$-almost all $\sigma$. Since it even converges for 
$K={\bf 1}$ expression (\ref{HolSum}) defines a bounded positive functional
on the operators $K$. We can find a unique positive trace-class operator
$\rho_\sigma$ such that
\[
tr(K \rho_\sigma)=\int_m \sum_k \langle A_k^\sigma \psi | K A_k^\sigma \psi \rangle
\]
for all compact $K$. The proof in \cite{HolevoRa} states furthermore
that $\cD$ can be the finite span of any orthogonal system of $\cH$.
Summarizing these results, we have:

\begin{Theorem}{\bf (Radon-Nikodym derivative of the instrument)} 
\label{Radon}

\noindent
For every orthonormal system $(|x_j\rangle)$ of $\cH$ there
is a family of CP-maps $G_\sigma$ defined on the finite 
span of all rank-one operators  $|x_i\rangle \langle x_j|$ 
such that
\[
G_m(\rho) =\int_m G_\sigma(\rho) d\nu(\sigma)
\]
for all $\rho$ in the domain of $G_\sigma$
and an appropriate $\sigma$-finite measure $\nu$.
\end{Theorem}

However,
the domain of the maps $G_\sigma$ can be extended:

\begin{Lemma}{\bf (Extended Domains)}\label{DomainExt}

Let $\rho\in T$ be arbitrary  and $\nu$ ne an arbitrary measure.
If the trace-class operators
$G_\sigma(\rho)$ are
consistently defined (in the sense that they define the density 
of the  measure $m \mapsto G_m(\rho)$) 
then we have:

\begin{enumerate}

\item The domain of $G_\sigma$ can consistently be extended 
to $\rho e^{-iHt}$ by
\[
G_\sigma(\rho e^{-iHt}):=G_\sigma(\rho) e^{-iHt} e^{i\sigma t}\,.
\]
Consistency means that it is the density of 
the measure $m\mapsto G_m(\rho e^{-iHt})$.
Similarly, we may define
$G_\sigma(\exp(iHt)\rho) :=\exp(-i\sigma t) \exp(iHt) G_\sigma(\rho)$.

\item 
Let $l$ be a measurable subset of $\R$ and $P_l$ be the projection onto
the space of functions vanishing on the complement of $l$.
Then one 
may extent the domain of $G_\sigma$ 
consistently to $\rho P_l$ by setting
\[
G_\sigma(\rho P_l):=G_\sigma (\rho) P_{l+\sigma}\,,
\]
and similarly $G_\sigma(P_l\rho):=P_{l+\sigma} G_\sigma(\rho)$.

\end{enumerate}

\end{Lemma}

\vspace{0.5cm}
\noindent
{\it Proof:} 
The Fourier transform of the measure
\[
m\mapsto tr (KG_m(\rho e^{-iHt}))
\]
is given by
\begin{eqnarray}
f(r)&:=&tr(K \hat{G}_r (\rho e^{-iHt}))=tr(K \hat{G}_{t+r}(\rho) \, e^{-iHt})\\
&=&\int tr(K G_\sigma(\rho) \,e^{-iHt}) e^{i(r+t)\sigma} d\nu(\sigma)\,. 
\label{Ext}
\end{eqnarray}
The last equality holds since $\hat{G}_{r+t}$ is the Fourier transform
of $m \mapsto G_m$ evaluated at $r+t$.
Set $\rho_\sigma:=G_\sigma(\rho)$.
We write (\ref{Ext}) as
\[
\int tr(K \rho_\sigma e^{-iHt} e^{it\sigma}) e^{ir\sigma} d\nu(\sigma)\,,
\]
which is the Fourier transform of a measure with density
\[
\sigma \mapsto tr(K \rho_\sigma \,e^{-iHt} e^{it\sigma}) \,.
\]
This proves statement (1).
  
To prove (2) we have to show that
\begin{equation}\label{Vergl}
tr(K G_m(\rho P_l))=\int_m G_\sigma (\rho)  P_{l+\sigma} d\nu(\sigma)
\end{equation}
holds for all $K$ and measurable $m$.
Consider the scalar complex-valued measure 
\[
l \mapsto tr(K G_m(\rho P_l))\,.
\]
Its Fourier transform is 
\[
f(t):= tr(K G_m( \rho e^{iHt}))=
\int_m tr(K \rho_\sigma e^{-iHt} e^{it\sigma}) d\nu(\sigma)\,,
\]
where the last equality is due to statement (1).
By 
\[
e^{-iHt} e^{it\sigma}=e^{i (H+\sigma {\bf 1})t}
\]
the last expression is the Fourier transform of the measure
\[
l \mapsto \int_m G_\sigma (\rho)  P_{l+\sigma}\, d\nu(\sigma)\,.
\]
This proves that both sides of eq.~(\ref{Vergl}) coincide.
$\Box$
\vspace{0.5cm}

We would like to characterize the maps $G_\sigma$ more explicitly.
We already have done this in Theorem~\ref{FiniteMain} for the finite 
dimensional case when the spectrum of $H$ is non-degenerate.
Now we 
assume that $\cH$ is the set of square integrable functions on $\R$
\[
\cH:=L^2(\R,\mu) \,,
\]
where $\mu$ is an arbitrary
measure on $\R$ defined on the Lebesgue measurable sets.
To formalize the assumption of non-degenerate spectrum
we define  $H$ as the multiplication operator
\[
(H\psi)(\omega)=\omega \psi(\omega) \,\,\,\forall \omega \in \R\,.
\]
Note that the Hadamard product can be generalized to
infinite dimensions since every density operator $\rho$  has a
representation $\rho=\sum_j p_j |\phi_j \rangle \langle \phi_j|$ with eigenvectors $|\phi_j\rangle$. Then we may interpret 
\[
\rho(\omega,\omega'):=\sum_j p_j \phi_j(\omega) \overline{\phi}_j (\omega')
\]
as the entries of $\rho$.  This is also the representation of
$\rho$ by its integral kernel, i.e., 
\[
(\rho\psi)(\omega)= 
\int \rho(\omega,\omega') \psi(\omega') d\mu(\omega')\,.
\]
Accordingly, we define:

\begin{Definition}{\bf (Hadamard product in infinite dimensions)}

\noindent
 For two trace-class operators $\rho,\rho'$ 
and a function $M: \R\times\R \rightarrow {\bf C}$
we write
\[
\rho'=M * \rho
\] 
if 
\[
\rho'(\omega,\omega')=M(\omega,\omega') \rho(\omega,\omega')\,.
\]
\end{Definition}

To generalize the concept of partial shifts to $L^2(\R,\mu)$ is
less straightforward.
Since they should be partial isometries,
the problem is two-fold:
Assume first that $\mu$ is given by a non-constant and non-zero
density  with respect to
the Lebesgue measure. Then it is intuitively 
clear that the generalized ``shift'' rescales the function
to compensate the different densities at different points
in order to be isometric.
Assume secondly that
$H$ has discrete and continuous spectrum, i.e.,
$\mu$ consists of singular and absolutely 
continuous parts with respect to the Lebesgue measure. 
Then one cannot expect that the discrete part
of the wave function can be shifted to the continuous part 
and vice versa. The part of the wave function where 
this is the case has to be mapped to zero.

Despite of these difficulties, 
the following lemma shows that our concept makes sense
even in the general situation:

\begin{Lemma}[Partial shifts]\label{PShifts}${}$
For every $\sigma \in \R$ 
there is a unique  partial isometry $S_\sigma$ with the following properties:

\begin{enumerate}
\item 
For all $f \in \cH$ we have
\[
(S_\sigma f)(\omega):=s_\sigma(\omega-\sigma)f(\omega-\sigma)\,,
\] 
where $s_\sigma$ is an appropriate
measurable ``scaling'' function.

\item Under all partial isometries $\tilde{S}_\sigma$ which have a representation as in (1) with an appropriate scaling function $\tilde{s}_\sigma$ 
the operator $S_\sigma$ has minimal kernel, i.e., its kernel is contained
in the kernel of all those $\tilde{S}_\sigma$.

\end{enumerate}

\end{Lemma}

{\it Proof:} Define the translated measure 
$\mu_\sigma(m):=\mu(m+\sigma)$.
Then there is a unique decomposition 
\[
\mu=\mu_c + \mu_s
\]
where $\mu_c$ is absolutely continuous with respect to $\mu_\sigma$
and $\mu_s$ is orthogonal to $\mu_\sigma$ \cite{Rao}.
Here, orthogonality  
means that there is a measurable set $B$ such that 
$\mu_s (B)=0$ and $\mu_\sigma (\overline{B})=0$ where $\overline{B}$ 
denotes the complement of $B$.
Let $p$ be the density of $\mu_c$ with respect to $\mu_\sigma$.
Set $s_\sigma:=\sqrt{p}$. 
First we show that $S_\sigma$ is an isometry of the subspace
$\cH_c$ given by all functions $f$ with
$f(\omega)=0$ for $\omega \in \overline{B}$:
\begin{eqnarray}\label{PartialM}
\int |s_\sigma(\omega-\sigma) f(\omega-\sigma)|^2 d\mu(\omega)
&=&\int |s_\sigma (\omega)|^2 |f(\omega)|^2 d\mu_\sigma(\omega)\\
=\int |f(\omega)|^2 p(\omega) d\mu_\sigma (\omega)
&=&\int |f(\omega)|^2 d\mu_c \nonumber
\end{eqnarray} 
This integral is equal to $\int |f(\omega)|^2 d\mu$ if
$f$ vanishes on $\overline{B}$.
On the other hand, it is zero if $f$ vanishes on $B$. 
This shows that the space can be decomposed
into the kernel of $S_\sigma$ and a  
subspace where $S_\sigma$ is an isometry. 
Hence we have $\|S_\sigma f\|\leq \|f\|$  on the whole space. 
Note that the equations showing this inequality show
furthermore that $S_\sigma$ is well-defined 
with respect to the
equivalence classes in $L^2(\R,\mu)$.

Let $\tilde{S}_\sigma$ be another isometry satisfying
condition (1) with the scaling function $\tilde{s}_\sigma$. 
Set $D:=\tilde{s}^{-1}(0)$. 
Clearly, condition (1) implies $[H,\tilde{S}_\sigma]=\sigma \tilde{S}_\sigma$ and 
$[H,\tilde{S}^\dagger_\sigma]=-\sigma \tilde{S}^\dagger_\sigma$.
 Hence the initial projection $\tilde{S}_\sigma^\dagger \tilde{S}_\sigma$ 
commutes with $H$ and is therefore the multiplication 
operator with the characteristic function $\chi_D$
of a measurable set $D$. Let $f$ vanish on $\overline{D}$. 
Since $\tilde{S}_\sigma$ 
preserves its norm we have
\[
\int |f(\omega)|^2 d\mu(\omega) =\int |f(\omega)|^2 
\tilde{s}^2_\sigma (\omega) d \mu_\sigma (\omega) \,.
\]
Since this holds for all functions which vanish on $\overline{D}$ 
$\mu$-almost everywhere it shows that the restriction of $\mu$ to $D$
is absolutely continuous with respect to $\mu_\sigma$ (with the density
$\tilde{s}_\sigma^2$). 
Hence the singular part of $\mu$ (with respect to $\mu_\sigma$) vanishes
on $D$ and we have
\[
\int |f(\omega)|^2 d\mu (\omega) = \int  |f(\omega)|^2 d\mu_c (\omega)\,.
\]
As can be seen in eq.~(\ref{PartialM}), this shows that
the norm of $f$ is also preserved
by $S_\sigma$. Hence the isometric subspace of $\tilde{S}_\sigma$ is a 
subspace of the isometric subspace of $S_\sigma$ and the kernel of
$S_\sigma$ is a subspace of the kernel of $\tilde{S}_\sigma$.$\Box$

\vspace{0.5cm}

Fortunately, these shifts have the following 
property:
\begin{Lemma}\label{Adjun}
The partial shift of Lemma \ref{PShifts} satisfy
\[
S_\sigma^\dagger=S_{-\sigma}\,.
\]
\end{Lemma}

{\it Proof:}
Rewriting the inner product of $\cH$ as an integral
one checks easily that $S_\sigma^\dagger$ has to be a translation
by $-\sigma$ with an appropriate scaling function.  
It is easy to see that $S_{-\sigma}S_\sigma =S_\sigma^\dagger S_\sigma$:
On the kernel of $S_\sigma$ both products coincide trivially. On its
orthogonal complement, i.e., the isometric subscape of
$S_\sigma$ the operator $S_\sigma^\dagger S_\sigma$ is the identity.
But this must also be true for $S_{-\sigma} S_\sigma$ because the isometric
subspace of $S_{-\sigma}$ must contain the isometric
subspace of $S_\sigma^\dagger$  since $S_{-\sigma}$ has minimal kernel
in the set of all partially isometric translations  
(see property (2) in Lemma \ref{PShifts}).
Using general properties of partial isometries
the image of $S_\sigma^\dagger$ satisfies therefore
\[
Im S_\sigma^\dagger = Im (S_\sigma^\dagger S_\sigma) = Im (S_{-\sigma} S_\sigma)
\leq Im S_{-\sigma}\,.
\]
Reversing this inequality by taking the orthogonal complement we
obtain
\[
ker S_\sigma \geq ker S_{-\sigma}^\dagger\,.
\]
Because this holds for all $\sigma \in \R$ we have
\[
ker S_{-\sigma} \geq ker S_\sigma^\dagger\,.
\]
Since $S_{-\sigma}$ has minimal kernel we conclude
\[
ker S_{-\sigma} = ker S_\sigma^\dagger\,.
\]
Due to the uniqueness of the maximal element in the set of
all partially isometric translations we have
\[
S_\sigma^\dagger =S_{-\sigma}\,.
\]
$\Box$

Note that we can only write $G_\sigma (\rho)$ in the form
$S_\sigma (M_\sigma *\rho)S_\sigma^\dagger$ if the density operator 
$G_\sigma(\rho)$ 
acts only on the image of $S_\sigma$ (which coincides with the image of 
the projection $S_\sigma S_\sigma^\dagger$).
But this is indeed the case:

\begin{Lemma}\label{KernelS}
If $\rho$ is in the domain of $G_\sigma$ its output
$\rho_\sigma:=G_\sigma(\rho)$ satisfies
\[
S_\sigma S_\sigma^\dagger \rho_\sigma S_\sigma S_\sigma^\dagger =\rho_\sigma\,.
\]
\end{Lemma}

\vspace{0.5cm}
\noindent
{\it Proof:}
We will show $G_\sigma(\rho)S_\sigma S_\sigma^\dagger =G_\sigma(\rho)$.
The statement $S_\sigma S_\sigma^\dagger G_\sigma(\rho)=G_\sigma(\rho)$ 
follows similarly.
Due to Lemma \ref{Adjun} we have
$S_\sigma S_\sigma^\dagger=S^\dagger_{-\sigma} S_{-\sigma}$, which
is the  projection onto the isometric subspace of
$S_{-\sigma}$. It can explictly be given as follows (see proof of Lemma
\ref{PShifts}): Let $\mu:=\mu_c\oplus \mu_s$ be the decomposition into
the absolutely continuous and singular part of $\mu$ with respect to
$\mu_{-\sigma}$.  Let $C$ be a set such that 
$\mu_{-\sigma}(\overline{C})=0$ and
$\mu_s(C)=0$. Then $S^\dagger_{-\sigma} S_{-\sigma}=P_C$.  
Due to $\mu(\overline{C}-\sigma)=\mu_{-\sigma} (\overline{C})=0$
we have 
$P_{\overline{C}-\sigma}=0$. 
This implies
\[
G_\sigma(\rho) P_{\overline{C}} =G_\sigma (\rho P_{\overline{C}-\sigma})=0\,.
\]
$\Box$
\vspace{0.5cm}

For the construction in the sequel we choose the ONS
in Theorem \ref{Radon} such that
its finite span contains a vector $|\psi\rangle$ with $\psi(\omega)\neq 0$ for
$\mu$-almost all $\omega$. 

Now we can define a function
\[
M_\sigma : \R \times \R \rightarrow {\bf C}
\]
by
\[
M_\sigma(\omega,\omega'):=\frac{k(\omega,\omega')}{\psi(\omega)\overline{\psi}(\omega')}\,,
\]
where the function $k$ 
represents 
the trace-class operator $S_\sigma^\dagger \rho_\sigma S_\sigma$.
Even though this function is not well-defined,
the 
calculations below show that the freedom of choosing $M_\sigma$ is
irrelevant.

By construction and using Lemma~\ref{KernelS}, we have 
\[
G_\sigma(\rho)=\rho_\sigma = S_\sigma (M_\sigma * \rho_\sigma) S_\sigma^\dagger\,.
\]
We conclude 
\[
G(|\psi \rangle \langle \psi |)=\int S_\sigma (M_\sigma * 
(|\psi\rangle \langle \psi|)) S_\sigma^\dagger d\nu(\sigma)
\]
with $|\psi\rangle$ as above.
Using Lemma~\ref{DomainExt}, part (1), we have
\[
\int G_\sigma( e^{-iHt}|\psi \rangle \langle \psi|
e^{iHs})\,d\nu(\sigma) = 
\int S_\sigma ( M_\sigma *  (e^{-iHt}|\psi \rangle \langle \psi|
e^{iHs})) S_\sigma^\dagger\,d\nu(\sigma)\,,
\]
for all $s,t\in \R$.
The finite span of these rank-one operators
is dense in $T$ since $H$ is the multiplication
with the identity and the span of all $\exp(-iHt)|\psi\rangle $  
is therefore dense in $\cH$.
We conclude:

\begin{Theorem}{\bf (Dephasing - Energy Shift Representation)} \label{Haupt}

\noindent
There is a family of functions
$(M_\sigma)_{\sigma \in \R}$ 
\[
M_\sigma :\R \times \R \rightarrow {\bf C}\,,
\]
and a $\sigma$-finite measure $\nu$ on $\R$ such that
for a dense set of density operators $\rho$
the decompostion
\[
G(\rho)=\int S_\sigma (M_\sigma * \rho) S^\dagger_\sigma \, d\nu(\sigma)
\]
holds.
\end{Theorem}

\vspace{0.5cm}

We would like to have an analogy to the statement that $M_\sigma$ 
is in the finite dimensional case a {\it positive} matrix.
However, 
we have specified $M_\sigma$ only as a function and not as an operator.
It seems straightforward to consider it as an operator
by
\[
(M_\sigma \psi)(\omega) :=\int M_\sigma(\omega,\omega')\psi(\omega') d\mu(\omega')\,.
\]
But it is easy to see that this is not in general well-defined.
Let, for instance, $G$ be the identity on the density operators
on $l^2(\Z)$. 
Then 
the decomposition of $G$ reduces to  $G(\rho)=M_0*\rho$ with
$M_0(\omega,\omega')=1$. The formal matrix multiplication
of this ``all-one'' matrix with any square-integrable sequence 
leads to infinite coefficients.
However, formally it is like a positive operator in the following sense.
\[
M_\sigma *(|\psi \rangle \langle \psi|)
\]
is by construction the trace-class operator $\rho_\sigma$.
Translated to finite dimensions this expression to
\[
D M_\sigma D^\dagger
\]
where $D$ is a non-singular diagonal matrix (in analogy to the
statement that $\psi(\omega)$ vanishes almost nowhere.
Such a matrix can only be positive if $M_\sigma$ is itself positive.
In this sense we consider $M_\sigma$ as positive even though it is 
not an operator.
In analogy to the finite dimensional case,
an exact measurement of
the energy of the environment (before and after the interaction has 
taken place)  would lead to a map
\[
S_\sigma (M_\sigma * \rho) S^\dagger_\sigma\,,
\]
which is only a dephasing channel up to the known energy shift.

It should be noted that (depending on $\nu$) the integral may
represent a finite, or countable infinite sum, or  a continuous
integral even though the Kraus representation of $G$ 
is always possible with a countable sum \cite{Kraus}.

\section{Superpositions between states in the same orbit}

\label{Bounds}

The ``Hadamard-channel'' $\rho \mapsto M*\rho$ with a positive matrix $M$ 
allows perfect classical information transfer by using energy eigenstates
as logical states. Its quantum capacity depends on the destruction
of off-diagonal elements, i.e., whether the output states are 
more or less stationary states. Explicitly one has the following 
lower bound on the quantum capacity:

\begin{Lemma}\label{HQC}
In dimension $n$,
the quantum capacity $Q$ of the 
channel $\rho \mapsto M*\rho$ with $M(\omega,\omega)=1$ for all 
$\omega\in \Omega$ satisfies
\[
Q\geq \log_2 (n) - S(M/n)
\]
where $S(.)$ denotes the von-Neumann entropy.
\end{Lemma}

{\it Proof:}
Let 
\begin{equation}\label{MZer}
M:= \sum_j |m_j\rangle \langle m_j|
\end{equation}
be a decomposition into mutually orthogonal rank-one operators
with non-normalized vectors $|m_j\rangle$. 
Let $D_j$ be the diagonal matrices having the coefficients of $|m_j\rangle$ 
as entries. 
Then we have
\[
G(\rho) =\sum_j  D_j \rho D^\dagger_j\,.
\]
The quantum capacity is given by the maximum of the coherent information
\cite{Devetak}.
Let $|\phi\rangle \langle \phi |$ a state on $\C^n \otimes \C^n$
and $\rho$ its restriction to the right component.
Then the coherent information is defined as  
\[
S(G(\rho))- S (id \otimes G) (|\phi \rangle \langle \phi|) \,,
\]
where $S$ denotes the von-Neumann entropy. 
Consider the maximally entangled state
\[
|\phi\rangle :=\frac{1}{\sqrt{n}} \sum_j |j\rangle \otimes |j\rangle\,.
\]
Note that $\rho$ is the maximally mixed state and that 
it is preserved by the channel.
Hence we have  $G(S(\rho))=\log_2(n)$ and
\begin{equation}\label{Zer}
(id\otimes G) (|\phi\rangle \langle \phi|) =\sum_j 
(1\otimes D_j) |\phi \rangle 
\langle \phi | (1\otimes D_j^\dagger) \,.
\end{equation}
Furthermore we have 
\[
\langle \phi | (1\otimes D^\dagger_j)  (1\otimes D_i) |\phi\rangle=
tr(D^\dagger_j D_i)=\langle m_j|m_i\rangle =0\,.
\]
Therefore the sum (\ref{Zer}) is already the spectral decomposition
of $(id\otimes G)(|\phi\rangle \langle \phi|)$. The eigenvalues are the square
of the length of each vector $(1\otimes D_j) |\phi\rangle$.
It is given by $\langle m_j|m_j\rangle/n$. 
Hence the entropy of the output state is the von-Neumann entropy
$S(M/n)$. 
$\Box$

\vspace{0.5cm}

Here the destruction of quantum information coincides with
the destruction of timing information in the sense that 
states on the same orbit of $\alpha_t$ become less distinguishable.
It is obvious that decoherence with respect to the energy eigenbasis
leads destroys always {\it timing} information in this sense.
Here we want to address the question what happens when 
one considers orthogonal states on the same orbit of the time evolution
as reference basis. Is it possible that a time-covariant
channel destroys superpositions between them without affecting the
basis states?
Consider two vector states $|\phi_0\rangle$ and 
$|\phi_s\rangle:=\exp(-iHs)|\phi_0\rangle$. 
Clearly,
$G(|\phi_0\rangle \langle \phi_0|)=|\phi_0\rangle\langle \phi_0|$
implies 
\[
 S_\sigma( M_\sigma *(|\phi_0\rangle\langle \phi_0|))S_\sigma^\dagger=0
\]
for all $\sigma\neq 0$. This follows from the observation that
the output state can only be pure if all output operators are linearly
dependent. Every  $\sigma \neq 0$ would  
lead to a different output state.
Furthermore we have 
\[
M_0*(|\phi_0\rangle \langle \phi_s|)=M_0*(|\phi_0\rangle\langle \phi_0|)
e^{iHs}=|\phi_0\rangle \langle \phi_s|\,.
\]
One can conclude easily that every density operator acting
on the subspace spanned by  $|\phi_0\rangle$ and $|\phi_s\rangle$ 
is preserved by $G$.
A little bit more general, one has:

\begin{Theorem}
Let $G$ be a time-covariant CP-map on a system with \newline non-degenerated 
Hamiltonian.
If there is any pure state $|\phi\rangle \langle \phi|$ with
\[
G(|\phi\rangle \langle \phi|)=|\phi\rangle \langle \phi|
\]
 then
$G$ leaves all density operators invariant which act on the
Hilbert subspace spanned by all $\exp(-iHt)|\phi\rangle$.
\end{Theorem}

The essential argument above is that there exist no  
states which are invariant with respect to $S_\sigma$ for some 
$\sigma\neq 0$.
An analogue statement would not be true for covariance 
with respect to a discretized time evolution:

Consider the Hilbert space $l^2(\Z)$ of square summable
two-sided sequences and the discrete time evolution given by
the translation $(U\psi)(n):=\psi(n+1)$.  Let $(|e_n\rangle)_{n\in \Z}$ 
be the canaonical basis of $l^2(\Z)$.
Then the channel
\[
\rho \mapsto \sum_{n \in \Z} 
|e_n\rangle \langle e_n| \rho |e_n\rangle \langle e_n|
\]
leaves all basis states $|e_n\rangle$ invariant but destroys all 
superpositions.

The assumption that a pure state is preserved by $G$ is rather strong.
Actually we want to figure out whether {\it distiguishability} of
different states in the same orbit may be conserved even though
their {\it superpositions}  are  distroyed. We consider the 
following extreme case:

\begin{Definition}[Reliable timing]${}$\\ \label{Relia}
A time-covariant channel $G$ has the ``reliable timing property''
(with respect to the time $s$)
if there exists an input density operator $\rho$ and a real number $s$ 
such that 
for $\rho_s:=\alpha_s(\rho)$ 
the outputs $G(\rho)$ and $G(\rho_s)$ are perfectly distinguishable,
i.e., the density matrices are mutually orthogonal.
\end{Definition}
It is easy to verify that the input state $\rho$ can be chosen to be
pure.
To justify the definition, we show that
this property appears in the following situation:
Assume a sender, say Alice, wants to send a signal to a receiver, say Bob.
Assume furthermore that it should be guaranteed that Bob receives the
signal in the time interval  $[t_1,t_2]$. 
This requires that 
the physical state $\rho$ of the signal (when it is sent) 
is perfectly distinguishable from the time evolved state $\rho_s$ with
$s:=t_1-t_2$. 
This
is due to the fact that the following ``measurement'' distinguishes
between them: wait the time $t_1$ and ask Bob whether he
has already received the signal. 
If the medium between Alice and Bob modifies the signal, we may model this 
by a time-covariant operation $G$ which has clearly to preserve
the distinguishability of the states $\rho$ and $\rho_s$.

It is clear that reliable transfer of classical information
requires two states $\rho$ and $\gamma$ such that
$G(\rho)$ and $G(\gamma)$ are mutually orthogonal. 
The remark above
shows that there are situations in classical
information processing where two orthogonal output states
should exists {\it on the same orbit} since 
reliable timing requires this feature.
To see that non-trivial channels with this property exist
one may construct a channel of the form
\[
\rho \mapsto \sum_j p_j S_{\sigma_j} \rho S^\dagger_{\sigma_j}
\]
where the $\sigma_j$ are chosen such that 
$\langle \phi| S^\dagger_{\sigma_i} S_{\sigma_j} \phi\rangle =0$ 
for $i\neq j$. This condition is, for instance,  satisfied if the
minimal distance between the values $\sigma_j$  exceeds  the spectral 
bandwidth of the input state. Real physical  channels will
satisfy condition \ref{Relia} at most approximatively.

In order to consider the quantum capacity of reliable timing channels we
choose $\rho$ and $\rho_s= U_s \rho U_s^\dagger$ 
with $U_s:=\exp(-iHs)$ such that $\rho$ and $\rho_s$ are mutually orthogonal. 
For simplicity we assume that
the dynamical evolution of $G(\rho)$ is periodic.
We obtain
a set 
\[
G(\rho_0),G(\rho_s),G(\rho_{2s}),\dots,G(\rho_{s(N-1)})
\]
 of $N$ 
mutually perfectly distinguishable density matrices.
Let $P$ be the support of $G(\rho)$, i.e. the smallest projection with
$PG(\rho)=G(\rho)$. Then the projections $U_{sj} PU^\dagger_{sj}$ 
with $j=0,\dots,N-1$ are mutually orthogonal.

Now we consider the channel with input 
\[
\C^N\equiv  span \{|\phi_{sj}\rangle \}_{j=0,\dots N-1}\,.
\]
It is clear that there exist pure input states 
$|\psi_0\rangle, \dots, |\psi_{s(N-1)}\rangle$ 
for which the orthogonality of the output states is satisfied.
By sending either of these $N$ states  $G$  
can  transfer $\log_2 N$ bits of classical information.
We want to derive sufficient  conditions under which $G$ allows also 
to send superpositions between the chosen basis states.
For doing so, we restrict $G$ to a channel $K$ on $N\times N$-density
matrices as follows. 

The input is restricted to the span
of all $|\psi_{sj}\rangle$ with $j=0,\dots,N-1$. The corresponding output
space is 
\[
\cH_r:= \oplus_{j=0}^{N-1} U_{sj} P U_{sj}^\dagger \cH \,.
\]
We may consider this space as the tensor product 
\[
\cH_r =\C^N \otimes P\cH
\]
if we identify the spaces $U_{sj} P U_{sj}^\dagger \cH$ for $j\neq 0$ 
with $P\cH$ via arbitrary 
unitaries. It is straightforward and convenient to 
chose the isomorphisms $U_{sj}$ with $j=0,\dots,N-1$.
For any output density operator $\gamma$ acting on $\cH_r$ 
the entry corresponding to $|j\rangle \langle k|$ of its restriction to $\C^N$ is given by
\[
tr ( U_{sk} P U^\dagger_{sj} \gamma)\,.
\]
To determine the channel $K$ we
have to compute all values
\[
tr( U_{sk} P U^\dagger_{sj} G(|\phi_{sl} \rangle \langle \phi_{sm}|))\,,
\]
for $k,j,l,m \in \{0,\dots,N-1\}$.
Due to the reliable timing property we know that
each state 
$|\phi_{sj}\rangle  \langle \phi_{sj}|$ leads to the output state
$|j\rangle \langle j|$. Furthermore the operator $|\phi_{sl}\rangle
\langle \phi_{sm}|$ leads with certainty to a multiple of
$|l\rangle \langle m|$. 
Roughly speaking, the reason is  
that a CP-map
which maps the states $|m\rangle \langle m|$ and $|l\rangle \langle l|$ 
onto itself maps also $|l\rangle \langle m|$ onto multiples 
of itself. This could, for instance, 
be shown by reformulating the CP-map
as the restriction of an appropriate unitary map (see \cite{Paulsen}).  
Hence we have only to determine which factors the off-diagonal terms
obtain. 
Due to the symmetry of the channel with respect to time
translations $U_{sj}$ we have only to evaluate 
\[
v(j):= tr( U_{sj} P G(|\phi_{0} \rangle \langle \phi_{sj}|))
=tr( P G(|\phi_{0} \rangle \langle \phi_{sj}| U_{sj}))\,.
\]
Then $K$ is explicitly given by
\[
K(\sum_{jk} c_{jk}|j\rangle \langle k|)=  
\sum_{jk} c_{jk} v(j-k) |j\rangle \langle k|\,,
\] 
This shows that the  channel $K$ is a ``Hadamard channel'' 
even though we
have not chosen the energy states as reference basis but the states
$|\phi_{sj}\rangle$ instead. 
Furthermore,  $K$ is given by
Hadamard multiplication  with
a {\it circulant} matrix $V$ with entries $V_{ij}:=v(i-j)$.
In order to calculate $v$ we use the 
explicit form of $G$ according to 
Theorem \ref{Radon}.
Using Lemma \ref{DomainExt} this yields
\[
v(j)=\int tr( G_\sigma (|\phi_0\rangle \langle \phi_0|) 
e^{-i\sigma sj} \, d\nu (\sigma)=f_{|\phi_0\rangle \langle
 \phi_0|,{\bf 1}}(-sj)\,,
\]
i.e., the Fourier transform of 
the energy shift probability measure
$\nu_{|\phi_0\rangle \langle \phi_0|,{\bf 1}}$
evaluated at the points $-sj$.
The eigenvalues $q_0,\dots,q_{M-1}$ 
of $V/n$ are given by the inverse Fourier transform
of $v$:
\[
q_k:=\frac{1}{N} \sum_{j=0}^{N-1} v(j) e^{-ijk}
\]
Using Lemma \ref{HQC} we conclude:

\begin{Theorem}[Q-capacity of channels with reliable timing]${}$\\
Let $G$ be a time-covariant channel with reliable
timing property with respect to the time $s$ and the initial state
$|\phi_0\rangle$. 
Let the dynamics be periodic with period length $sN$.
 Then the quantum capacity
can  only be zero if the Fourier transform $\hat{p}$ of  the  
probability measure 
$p(m):=tr(G_m(|\phi_0\rangle \langle \phi_0|))$ 
has zeros at $sj$ for $j=1,\dots,N-1$.
Otherwise the quantum capacity is at least
\[
\log_2 N -\sum_k  q_k \log_2 q_k,,
\]
where $q=(q_0,\dots,q_{N-1})$ is the discrete Fourier transform 
of the evaluation of
$f_{|\phi_0\rangle \langle \phi_0|,{\bf 1}}$  
on the $N$ points $0,-s,-2s,\dots,-(N-1)s$.
\end{Theorem}

Of course there are many possibilities to define finite dimensional
channels from the original one.
However, 
to study whether $G$ destroys 
superpositions between different states in the orbit
(``different pointer states of a clock'')
one has always to consider
\[
G(|\phi_0\rangle \langle \phi_{sj}|)=\hat{G}_{-sj}(|\phi_0\rangle\langle\phi_0 |)
e^{iH sj} \,.
\]
This shows that the Fourier transform of the measure $(G_m)$ is
decisive. For deriving the lower bound above we have evaluated
it for the observable ${\bf 1}$ which leads
to the Fourier transform of the POVM $(Q_m)$ defined in the last section. 
However, ${\bf 1}$ is not necessarily optimal for detecting superpositions
between the output states.

\section{Gaussian channel on a single mode}

\label{Gauss}

The Hilbert space of a single mode in quantum optics is
$l^2(\N_0)$, the set of square summable sequences. 
It can also be interpreted as the energy levels
of a harmonic  oscillator. 
A time-covariant channel which is often considered
is the following. Using position and momentum observables
$X$ and $P$, respectively, we define a translation $(a_1,a_2)\in \R^2$ by
the unitary transformation $\exp(i(a_1 X + a_2P))$, where 
$X:=a^\dagger +a$ and $P:=(a^\dagger -a)/i$.
It is convenient to introduce variables $r\in \R_0^+$ and $z$ on the
complex unit circle
$\Gamma$ by $a_1+ia_2=rz$. 
Rewriting the 
translation with operators $a^\dagger$ and $a$ and the 
parameters $z,r$ 
we obtain the translation operator (compare  \cite{Davidovich}) by
\[
D(z,r):=\exp(r (\overline{z} a^\dagger - z a))=
e^{ir^2/2} e^{r\overline{z}a^\dagger} e^{-rz a}\,. 
\]
Since the global phase factor is irrelevant we will,
in abuse of notation, use $D(z,r)$ for the term without
this factor.

Let the gaussian channel be given by a random displacement
$rz$ according to the two-dimensional rotation invariant
gauss distribution. Note that rotation symmetry is necessary 
in order to obtain a time-covariant channel since the Hamiltonian
$H:=diag(0,1,2,\dots)$ 
corresponds to a rotation in the ``phase space''.
Note that the possible energy shifts are in $\Z$, hence
we expect a countable sum of $G_\sigma$ which are (in contrary to the
general case) not only densely defined.  

The whole channel is given by (see \cite{Giovannetti})
\[
G(\rho)= \int \int_\Gamma D(z,r) \rho D(\overline{z},r) \,dz\, p(r) dr
\]
with 
$p(r):=\exp(-r^2/(2s^2))r/s^2$ and $s$ denotes
the standard deviation.

As usual, we
introduce the creation operator $a^\dagger$ and the annihilation 
operator $a$ by $a|j\rangle:= \sqrt{j} |j-1\rangle$ for $j\geq 1$
and $a|0\rangle =0$.  
Then we
write $D(z,r)$ as the power series
\[
D(z,r)= e^{r\overline{z} a^\dagger}e^{-rza}
=  \sum_{m\geq 0} \frac{(r\overline{z})^n (a^\dagger) ^m}{m!}
\sum_{n\geq 0} \frac{(-rz)^n a^n}{n!}\,.
\]
It decomposes canonically into terms $D_\sigma(z,r)$ 
with $\sigma \in \Z$ 
satisfying the commutation relation $[H,D_\sigma(z,r)]=\sigma D_\sigma(z,r)$ 
if we define
\[
D_\sigma(z,r):= \sum_{n\geq -\sigma,0} 
\frac{(r\overline{z})^{n+\sigma} (a^\dagger) ^{n+\sigma}}{(n+\sigma)!}
\frac{(-rz)^n a^n}{n!}=
\overline{z}^\sigma r^\sigma  \sum_{n\geq -\sigma,0} (-1)^n 
 \frac{(a^\dagger) ^{n+\sigma}}{(n+\sigma)!}
\frac{a^n}{n!} (r^2)^n\,.
\]
Note that $D_\sigma(z,r)$ maps states with energy $j$ onto states
with energy $j+\sigma$. This suggests already that they may correspond
to the maps $G_\sigma$ of Section \ref{Form}.
The operator
\[
D_\sigma(r):=D_\sigma (z,r) z^{\sigma} \sqrt{2\pi}
\]
is independent of $z$. 
We 
conclude therefore that terms of the form $D_\sigma (\overline{z},r)
\rho D_{\sigma'} (z,r)$ cancel for $\sigma\neq \sigma'$ after 
integration over 
all $z\in \Gamma$. 
We obtain
\[
G(\rho)=\sum_{\sigma \in \Z}   \int  D_\sigma (r) 
\rho D_\sigma(r)  \,p(r) dr\,.
\]
Consider the case $\sigma\geq 0$ first.
Then we have
\begin{eqnarray*}
(a^\dagger)^{n+\sigma} a^n |j\rangle &=& 
\sqrt{(j+\sigma) \dots (j+1)}\,\, j(j-1) \dots (j-n+1) |j+\sigma \rangle \\&=&
\frac{1}{\sqrt{(j+1)\dots (j+\sigma)}}\,\,\,\frac{(j+\sigma)!}{(j-n)!} 
|j+\sigma\rangle\,,
\end{eqnarray*}
for all $n\leq j$.
For the case $\sigma <0$ we obtain
\begin{eqnarray*}
(a^\dagger)^{n+\sigma} a^n |j\rangle &=&
\sqrt{j(j-1)\dots (j+\sigma+1)}\,\,\,\frac{(j+\sigma)!}{(j-n)!} 
|j+\sigma\rangle\,,
\end{eqnarray*}
For $\sigma\geq 0$ we
conclude
\begin{eqnarray*}
D_\sigma(r)|j\rangle &=&
\frac{r^\sigma}{\sqrt{(j+1)\dots (j+\sigma)}}
\sum_{n=0}^j (-1)^n
\frac{(j+\sigma)!}{(j-n)!\, (n+\sigma)! \,n!} (r^2)^n  
|j+\sigma\rangle\\&=&
\frac{r^\sigma}{\sqrt{(j+1)\dots (j+\sigma)}}
L^\sigma_j (r^2) |j+\sigma\rangle\,,
\end{eqnarray*}
where
\[
L^{(\sigma)}_j(x):=
\sum_{n=0}^j (-1)^n \frac{(j+\sigma)!}{(j-n)!\,(n+\sigma)!\, n!}
(x) 
\]
is a Laguerre polynomial \cite{Buchholz}. 

Hence we have
\[
M_\sigma(j,j')=\int \frac{(r^2)^\sigma}{\sqrt{(j+1)\dots (j+\sigma)
\,\,(j'+1)\dots (j'+\sigma)}}
L^{(\sigma)}_{j} (r^2)
L^{(\sigma)}_{j'} (r^2) p(r) dr\,.
\]
For negative $\sigma$ we may express $M_\sigma$ similarly by Laguerre polynomials together with a different factor.
Hence the decomposition of the single mode 
gaussian channel can be given in a closed form even though it leads to
a less familiar representation.

\section{Conclusions}

We have shown that every time-covariant CP-map has 
a representation as an integral over a family of CP-maps. 
If the spectrum of the system Hamiltonian is non-degenerated,
each of these
components consists of a Hadamard multiplication
with a positive operator
followed by an energy shift.
Formally, the output of the channel is an unselected post-measurement
state.  The measured (and ignored) quantity is the energy which has been
transferred to or from the environment.
Conditional to the measurement outcome a different dephasing channel is 
applied.

Furthermore we have addressed the question to what extent covariant
channels can destroy superpositions between $N$ mutually orthogonal states
on the same orbit of the time evolution.
For a specific type of channels (with ``reliable timing property'')
we have shown 
that the general decomposition helps
to derive lower bounds on the quantum capacity.

The decomposition presented here may
be a helpful approach to describe a rather general type
of decoherence and relaxation phenomena.

We have calculated the decomposition  explicitly for 
a rotation invariant 
gaussian channel acting on the state space of a single mode
Fock space. The dephasing operations can then be described
using Laguerre polynomials.

\section*{Acknowledgments}

Thanks to J\"{u}rgen Schweizer for useful discussions and 
Leonid Gurvitz for a helpful remark.
This work has been supported by the DFG-project
``Komplexit\"{a}t und Energie'' in the SPP VIVA.
Part of this work has been done during a visit at Los Alamos National 
Laboratory.



\end{document}